\documentclass[]{spie}  

\usepackage{amsmath,amsfonts,amssymb}
\usepackage{graphicx}
\usepackage{subcaption}
\usepackage[colorlinks=true, allcolors=blue]{hyperref}
\usepackage{multirow}
\usepackage[perpage]{footmisc}

\title{Design, development, and commissioning of a flexible test setup for the AXIS prototype detector}

\author[a]{Abigail Y. Pan}
\author[a]{Haley R. Stueber}
\author[a]{Tanmoy Chattopadhyay}
\author[a]{Steven W. Allen}
\author[b]{Marshall W. Bautz}
\author[c]{Kevan Donlon}
\author[b]{Catherine E. Grant}
\author[a]{Sven Hermann}
\author[b]{Beverly LaMarr}
\author[b]{Andrew Malonis}
\author[b]{Eric D. Miller}
\author[a]{Glenn Morris}
\author[a]{Peter Orel}
\author[a]{Artem Poliszczuk}
\author[b]{Gregory Prigozhin}
\author[a]{Dan Wilkins}
\affil[a]{Kavli Institute for Particle Astrophysics and Cosmology, Stanford University, 452 Lomita Mall, Stanford, CA 94305, USA}
\affil[b]{MIT Kavli Institute for Astrophysics and Space Research, Massachusetts Institute of Technology, Cambridge, MA, USA}
\affil[c]{MIT Lincoln Laboratory, Lexington, MA, USA}

\authorinfo{Further author information: (Send correspondence to Abigail Y. Pan)\\Abigail Y. Pan: E-mail: apan5@stanford.edu}

\begin{document} 
\maketitle

\begin{abstract}
The Advanced X-ray Imaging Satellite (AXIS) is one of two candidate mission concepts selected for Phase-A study for the new NASA Astrophysics Probe Explorer (APEX) mission class, with a planned launch in 2032. The X-ray camera for AXIS is under joint development by the X-ray Astronomy and Observational Cosmology (XOC) Group at Stanford, the MIT Kavli Institute (MKI), and MIT Lincoln Laboratory (MIT-LL). To accelerate development efforts and meet the AXIS mission requirements, XOC has developed a twin beamline testing system, capable of providing the necessary performance, flexibility, and robustness. We present design details, simulations, and performance results for the newer of the two beamlines, constructed and optimized to test and characterize the first full-size MIT-LL AXIS prototype detectors, operating with the Stanford-developed Multi-Channel Readout Chip (MCRC) integrated readout electronics system. The XOC X-ray beamline design is forward-looking and flexible, with a modular structure adaptable to a wide range of detector technologies identified by the Great Observatories Maturation Program (GOMAP) that span the X-ray to near-infrared wavelengths. 
\end{abstract}


\section{INTRODUCTION}
\label{sec:intro}  
In response to the recommendations of the US National Academy of Sciences 2020 Decadal Survey of Astronomy and Astrophysics (Astro2020), NASA has created the Astrophysics Probe Explorer (APEX)\cite{national_aeronautics_and_space_administration_announcement_2023} line and Great Observatories Mission and Technology Maturation Program (GOMaP)\cite{julie_crooke_great_2023} to develop future space mission concepts. In its 2024 Prioritized Technology Gap Report, NASA identified large, fast, low noise, X-ray detectors as a Tier 1 gap essential to the operation of future Probe- and Great Observatory class X-ray missions\cite{noauthor_2024_2024}. The high spatial resolution and wide field-of-view of future X-ray observatories\cite{miller_high-speed_2023} will require large-format imaging arrays with small pixel sizes. Operation at high frame rates and with low readout noise will be essential to minimize the effects of dark current\cite{charles_r_proffitt_stis_2015}, the cosmic ray-induced particle background \cite{wilkins_augmenting_2024, poliszczuk_towards_2024} and photon pileup\cite{davis_event_2001, lumb_simulations_2000, mccollough_impact_2005}, and to achieve the required target sensitivities in the 0.3-10 keV energy range.

The X-ray Astronomy and Observational Cosmology (XOC) group at Stanford is working with the MIT Kavli Institute (MKI) and MIT Lincoln Laboratory (MIT-LL) to develop detectors and readout electronics to address this gap. The collaboration is jointly developing the X-ray camera for the the Advanced X-ray Imaging Satellite (AXIS)\cite{reynolds_overview_2023}, a Probe-class mission concept currently in Phase-A study for the APEX program. As part of this effort, we are designing and testing the representative flight hardware prototype for the AXIS focal plane, which includes a full-sized MIT-LL CCID-100 charge coupled device (CCD) detector coupled to two Multi-Channel Readout Chips (MCRC), a Stanford-designed application specific integrated circuit (ASIC) for fast and low-noise detector readout\cite{miller_high-speed_2023}.

In addition to work on AXIS, our collaboration is also developing other detector and readout technologies, including the Single electron Sensitive Read Out (SiSeRO) device\cite{herrmann_mcrc_2020}, a charge detection technology capable of producing sub-electron noise through repetitive non-destructive readout techniques \cite{bautz_toward_2018}. We are currently implementing SiSeROs as an on-chip output amplifier stage on the MIT-LL CCID-93 detector and its variants\cite{chattopadhyay_first_2022, chattopadhyay_demonstrating_2024}.

To support this array of development efforts, the XOC group has built a second detector and readout electronics test system, the XOC Gen2.0 X-ray Beamline. In this manuscript, we discuss the design and commissioning of this beamline, as well as the test setup for the MIT-CCID-100 detectors \cite{miller_high-speed_2023}. 

\section{The Second Generation XOC X-ray Beamline}

\begin{figure} [ht]
   \begin{center}
   \begin{tabular}{c} 
   \includegraphics[width=0.9\textwidth]{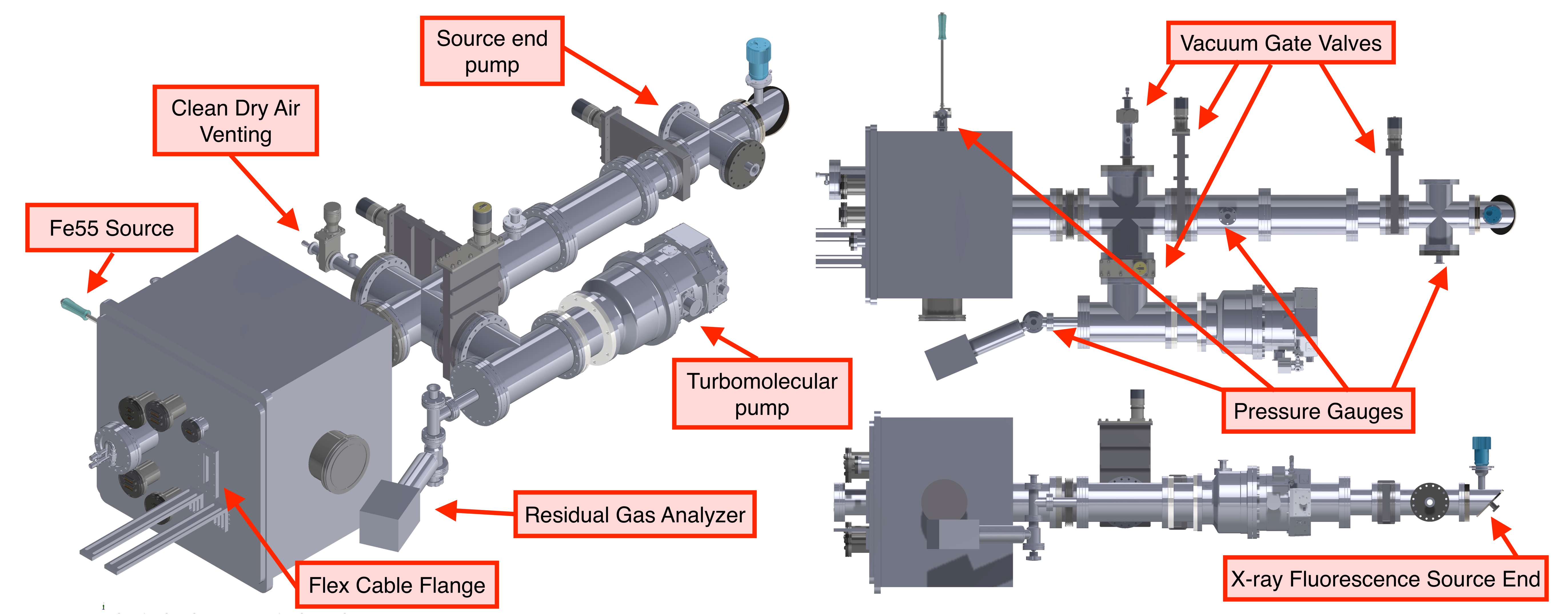}
   \end{tabular}
   \end{center}
   \caption[SolidworksMod] 
   { \label{fig:SolidworksMod} 
Isometric, top, and side view of the XOC Gen2.0 Beamline Solidworks model. The detector readout and cooling systems are mounted to the vacuum chamber door. Connection to an external Archon CCD controller is made through a vacuum potted flex cable. The chamber is also fitted with an Fe$^{55}$ source. Backfill during venting and purging is done with clean dry air. A residual gas analyzer is installed for contamination monitoring (\textbf{Left}). Four vacuum gate valves and pressure gauges are used for vacuum control and monitoring (\textbf{Top Right}). The X-ray fluorescence setup is located at the other end of the 2.5 m beamline. (\textbf{Bottom Right}). }
   \end{figure}

\begin{figure} [ht]
   \begin{center}
   \begin{tabular}{c} 
   \includegraphics[width=0.9\textwidth]{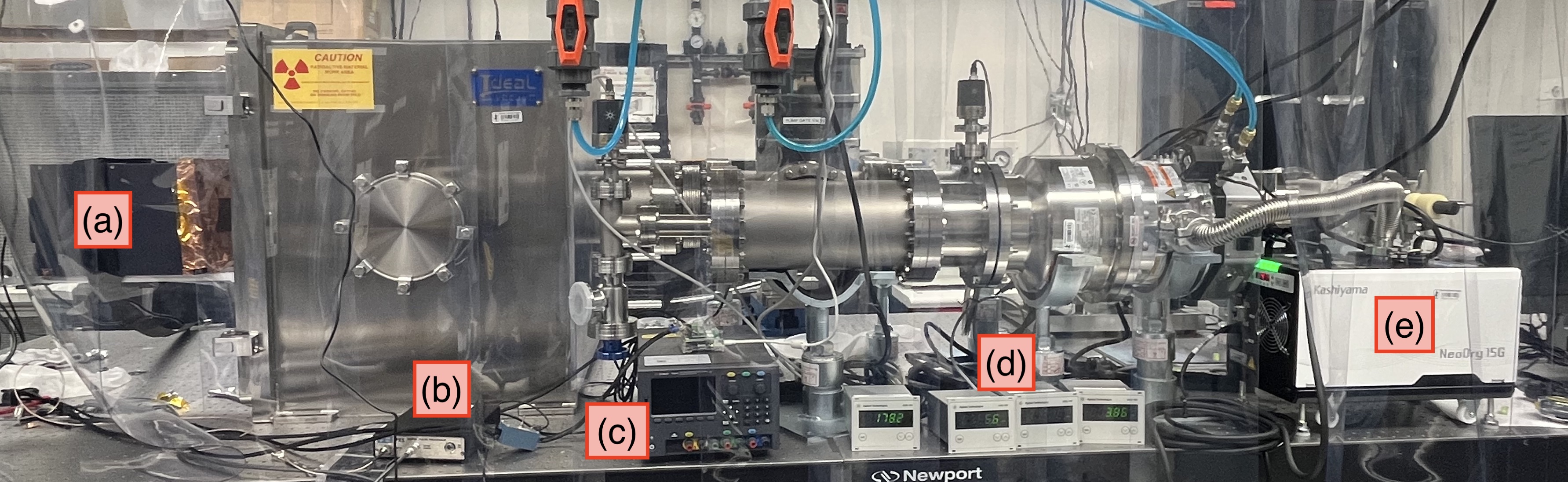}
   \end{tabular}
   \end{center}
   \caption[beamassemble] 
   { \label{fig:beamassemble} 
The assembled XOC Gen2.0 Beamline. The vacuum chamber can be seen on the left and the source end on the right. Additional electronic equipment has been labelled, including the Archon CCD controller (\textbf{a}), the SDD monitor digital pulse processor (\textbf{b}), temperature monitor and PID system (\textbf{c}), vacuum gauge displays (\textbf{d}) and roughing pump (\textbf{e}).     }
   \end{figure}

The XOC Gen2.0 X-ray Beamline is a 2.5 meter long  test system fitted with a vacuum detector chamber and an X-ray fluorescence (XRF) setup on either end, respectively. Similar to its predecessor, the Gen2.0 beamline is equipped with multiple vacuum gate valves, allowing the chamber and XRF source end to be independently brought to atmosphere while maintaining vacuum in the main body. Detectors and readout electronics are mounted to the vacuum chamber door using a modular mounting plate, providing the flexibility to support a range of detector characterization efforts. The XRF source end contains an interchangeable wheel of user-selected source materials capable of producing monoenergetic X-ray emission lines across the 0.3 - 10 keV energy range of interest. The Solidworks model of the beamline is shown in Fig. \ref{fig:SolidworksMod} and the fully assembled setup is shown in Fig. \ref{fig:beamassemble}. Based on the performance of the Gen1.0 Beamline\cite{stueber_xoc_2024}, we have redesigned the vacuum system, mechanical and cooling assembly, and source end for the Gen2.0 Beamline. This paper discusses these changes, as well as updates that have been made to accommodate the commissioning and characterization of the CCID-100 detector.

\section{Vacuum System}
\label{sec:Vac}
Two sets of vacuum pumps are installed in beamline, allowing the chamber- and source-side to be pumped down independently. We use the same Agilent TPS-flexy Turbo Pumping System\footnote{https://www.agilent.com/store/productDetail.jsp?catalogId=X1699-64087\&catId=SubCat2ECS\_1466018} configuration installed in the Gen1.0 beamline to pump out the source-side. The chamber side has been redesigned with a NeoDry-15G roughing root pump\footnote{https://www.kashiyama.com/en/pumps/product/neodry-g/} and a Shimadzu X1205 turbomolecular pump\footnote{https://www.shimadzu.com/industry/products/tmp/ittai/tmp1205.html}. The new chamber-side vacuums were selected to reduce operating noise and maintenance requirements, as well as increasing the efficiency of our pump-down and venting time\cite{svichkar_calculation_2018}. The system is able to reach \textless1e-6 mbar pressures in an hour, and the spin-down time of the turbo pump is roughly thirty minutes, allowing the system to be evacuated or brought to atmosphere more quickly. Control of both pumps can also be integrated into our remote lab monitoring systems through RS232 serial connections. 

The quality of our vacuum system is monitored by four Agilent FRG-700 Full Range Pirani Everted Magnetron pressure gauges \footnote{https://www.agilent.com/store/productDetail.jsp?catalogId=FRG700KF25\&catId=SubCat2ECS\_1465644} mounted along the length of the beamline (locations indicated in Fig. \ref{fig:SolidworksMod}). A Python monitoring script alerts lab members via Slack alerts if the pressure rises above 5e-5 mbar. As an additional check, we have installed a Kurt J. Lesker Faraday cup Residual Gas Analyzer (RGA)\footnote{https://www.lesker.com/residual-gas-analyzers-rga/kjlc-residual-gas-analyzers-rga/part/ele1fa000}, which acts as an in-situ leak and contamination monitor. The RGA is mounted with an isolation valve to reduce contamination when the unit is not in use, or when the detector chamber and beamline body are at atmosphere.

We utilize the same fail-safes implemented in the Gen1.0 Beamline to maintain personnel safety and the overall health of beamline components\cite{stueber_xoc_2024}. The chamber is vented and purged with clean dry air. A pressure relief valve\footnote{https://www.lesker.com/newweb/valves/vat/series.cfm?category=e\&series=212\#/series.cfm?category=e\&series=212} is designed to open when beamline pressure exceeds 1.2 atm, which prevents over-pressurization. In case of power outages, all of the vacuum pumps are run on uninterruptible power supplies (UPSs)\footnote{https://www.digikey.com/en/products/detail/tripp-lite/9PX2000RTN-L/17766589} \footnote{https://www.digikey.com/en/products/detail/tripp-lite/9PXEBM72RT-L/17766706}. This ensures that the turbo pumps remain under vacuum during operation, and prevents sudden re-pressurization damage to the XRF source or condensation buildup on the detector while cooling.

\section{Mechanical and Cooling Assembly}
\label{sec:Mech}

The detector and its electronic readout and cooling modules are mounted to the vacuum chamber door during testing. In order to test the detectors at representative operational temperatures, we utilize an Edwards Polycold PCC Compact Cryo-cooler\footnote{https://www.digikey.com/en/products/detail/laird-technologies-thermal-materials/A14556-01/2633557}, which is designed to achieve temperatures down to 70K. To accommodate the larger size and thermal mass of the CCID-100 detector package, we have updated the cooling and mechanical assembly. 
  \begin{figure} [ht]
   \begin{center}
   \begin{tabular}{c} 
   \includegraphics[width=0.9\textwidth]{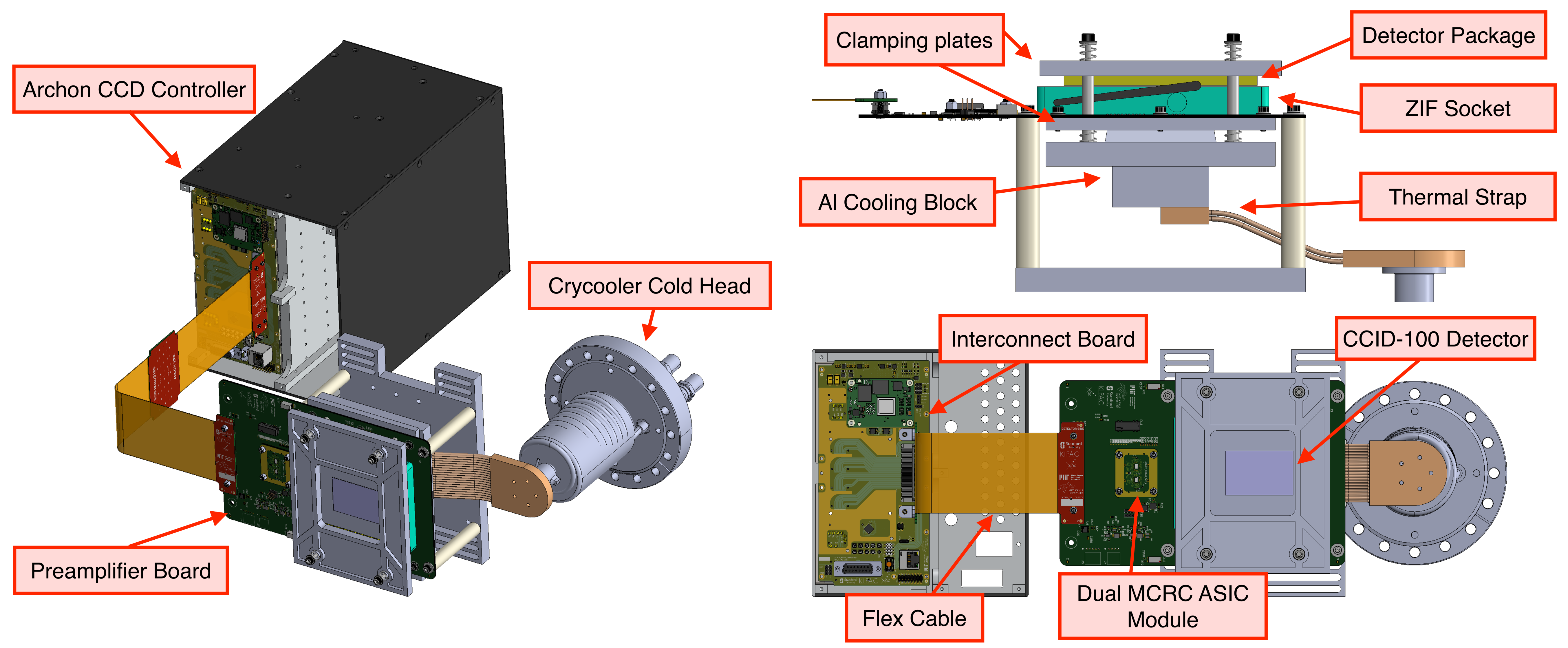}
   \end{tabular}
   \end{center}
   \caption[coolsetup] 
   { \label{fig:coolsetup} 
Solidworks model of the CCID-100 detector mechanical and cooling assembly and readout electronics chain. The detector and readout pre-amplifier board are mounted to the vacuum door. Connections to the CCD controller and cryocooler are done through a vacuum potted flex cable and the vacuum fitting compatible cryocooler cold head, respectively. A side profile of the mechanical assembly is pictured on the \textbf{top right}. Components of the readout electronics chain are labeled in the face-on view in the \textbf{bottom right}} 
   \end{figure}    
Thermal coupling between the cryocooler cold-head and the detector are now achieved with a customized copper strap manufactured by Technology Applications, Inc\footnote{https://www.techapps.com/}, and a larger aluminum block matching the size of the detector package. The detector is mounted to the pre-amplifier board using a zero insertion force (ZIF) socket and clamped to sit flush with the front face of the block. To ensure uniform thermal contact in the system, we add a layer of 100 $\mu$m indium foil between the copper strap and aluminum block, as well as a 508 $\mu$m layer of Tflex-500 thermally conductive gap filler padding\footnote{https://www.digikey.com/en/products/detail/laird-technologies-thermal-materials/A14556-01/2633557} between the block and detector. 

\begin{figure}
\centering
\begin{subfigure}{0.45\textwidth}
    \includegraphics[width=\textwidth]{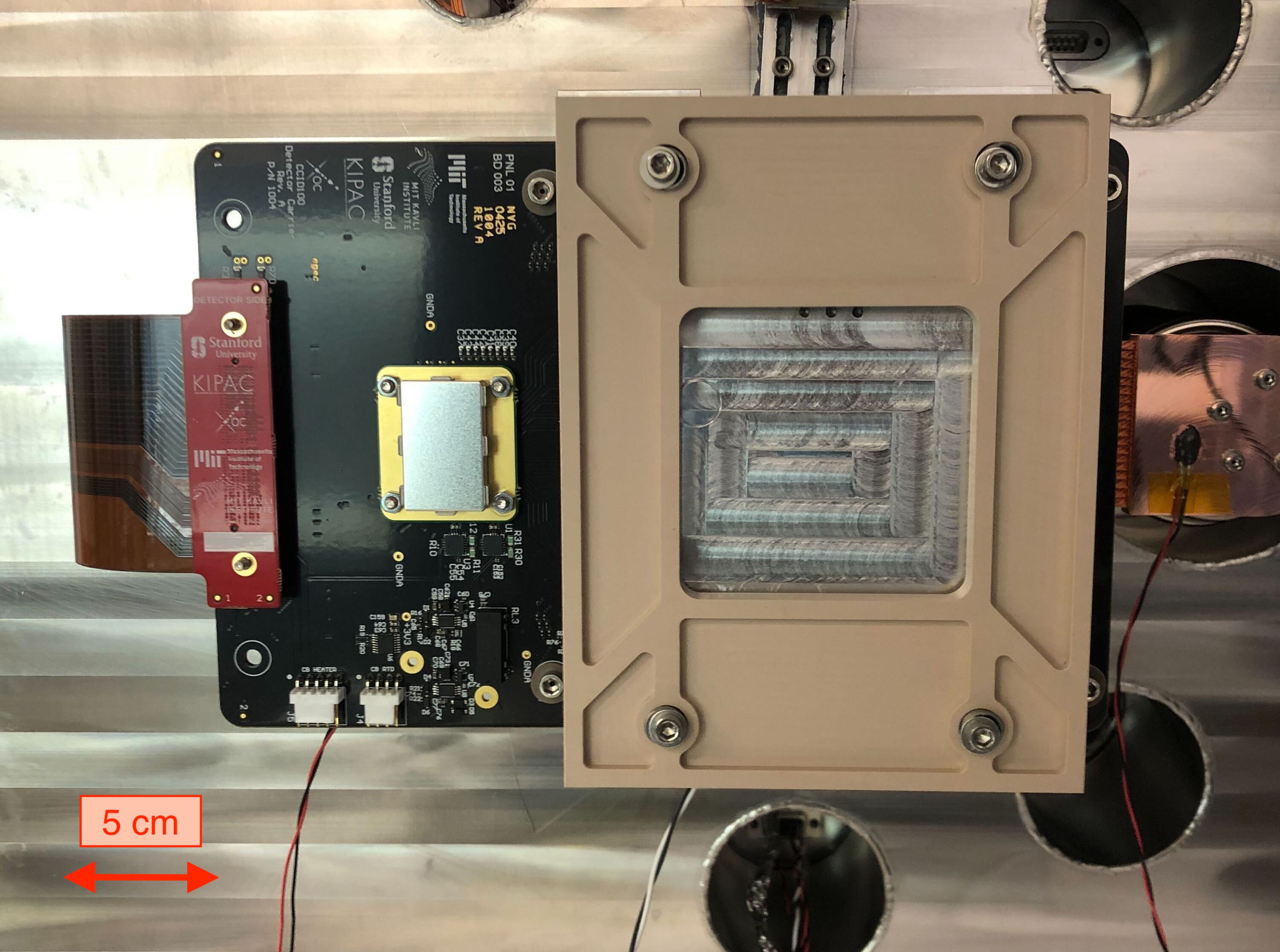}
    \caption{}
    \label{fig:coolingass_real}
\end{subfigure}
\hspace{1cm}
\begin{subfigure}{0.45\textwidth}
    \includegraphics[width=\textwidth]{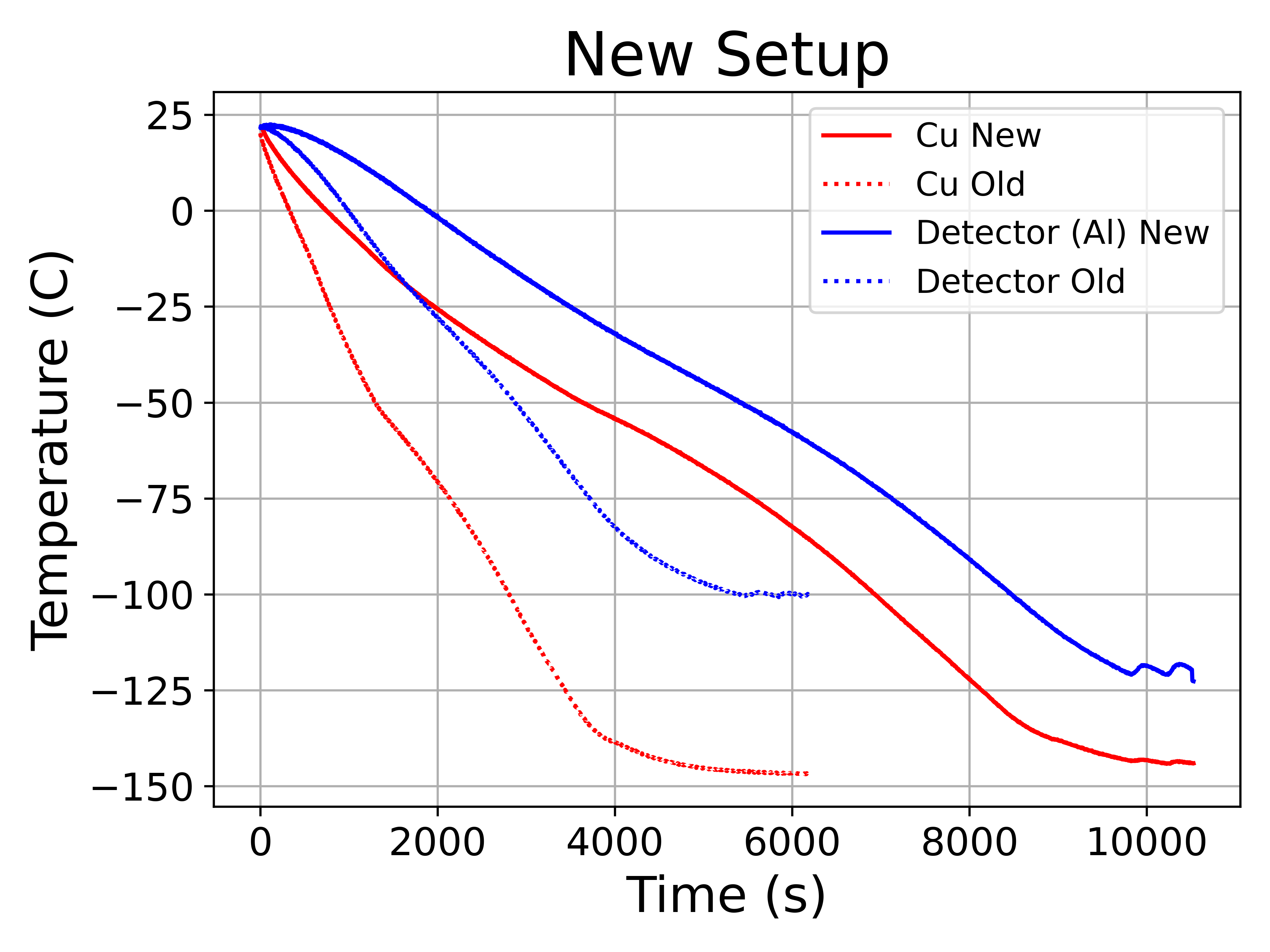}
    \caption{}
    \label{fig:coolingtest}
\end{subfigure}
\caption[cooling] 
{ \label{fig:cooling}(a) The fully assembled mechanical and cooling assembly. An aluminum block was used as a ``dummy" detector package for initial commissioning purposes. An RTD can be seen epoxied to the copper strap on the right side of the image. (b)  Performance comparison between the old (\textit{dashed line}) and new (\textit{solid line}) mechanical and cooling assembly. The new assembly was able to reach 20K cooler than previously achievable. Thermal losses between the copper strap and aluminum block have also been reduced. }
\end{figure}

We redesigned the mechanical assembly to improve thermal clamping and streamline the detector mounting process. The schematic for the assembly is shown in Fig. \ref{fig:coolsetup}. Rather than being installed separately, the aluminum cooling block is free-floating and spring clamped to the detector package. The springs are installed on four stainless steel stand-offs, which also align the block, detector package, and clamping plates. The use of spring clamping allows the assembly to maintain proper thermal contact as materials contract during the cooling process. The assembly is mounted to the door using with a second set of thermally isolated standoffs. The fully assembled system (with a ``dummy'' aluminum detector package for commissioning purposes) is shown in Fig. \ref{fig:coolingass_real}.

To control cooling rates and maintain stable operating temperatures for the detector, we employ a similar Python-based proportional-integral-derivative (PID) temperature control loop to the system installed in the Gen1.0 beamline\cite{wang_pid_2018}. The temperatures of the copper strap and aluminum block are monitored with 2-wire PT 1000 Tewa Sensor LLC resistance temperature detectors (RTDs)\footnote{https://www.digikey.com/en/products/detail/tewa-sensors-llc/TT-PT-1000B-2050-11-AUNI/9817197}, adhered to each component with a thermally conductive and electrically insulating epoxy\footnote{https://epoxyinternational.com/strongbond-53-thermally-conductive-electrically-insulating-compound-2-part-thixtropic-low-nasa-outgassing-adhesive}. The CCID-100 detector also has an on-chip temperature sensor, which is wired through the readout amplification board. The RTDs are monitored by a Raspberry Pi\footnote{https://www.raspberrypi.com/products/raspberry-pi-4-model-b/?variant=raspberry-pi-4-model-b-1gb}, which runs the digital PID temperature control loop\cite{noauthor_ivpid_nodate}.  We use a vacuum-rated 25 $\Omega$ Lakeshore heater cartridge\footnote{https://shop.lakeshore.com/default/heater-cartridge-0-25-in-diameter-25-ohm-100-whtr-25-100.html} embedded in the aluminum block as the PID actuator, which is powered by a Keysight E36313A DC Power Supply\footnote{https://www.keysight.com/us/en/product/E36313A/160w-triple-output-power-supply-6v-10a-2x-25v-2a.html} (See (c) in Fig. \ref{fig:beamassemble}). 
   
Initial cooling tests (Fig. \ref{fig:coolingtest}) demonstrate that the new assembly is able to achieve detector package temperatures of 153K (-120\textdegree{C}) in under 3 hours, 20K cooler than in our previous setup. Thermal loss across the system has also been reduced, with a 10K difference between the copper strap and aluminum block, and a roughly 3K loss between the block and an aluminum "dummy" detector package. 

\section{Electronic readout module}
\label{sec:Readout}

The basic electronics readout chain of the XOC beamline starts with a detector and custom pre-amplifier board mounted inside the vacuum chamber, which is connected to an external CCD controller through a vacuum potted flex cable (Fig. \ref{fig:coolsetup}). Detector output signals are amplified through the pre-amplifier board, which has two readout modes. The first mode is an analog signal chain composed of discrete electronic components. The second mode uses an application-specific integrated circuit (ASIC) chip called the Multi-Channel Readout Chip (MCRC) to amplify, buffer, and convert the single-ended signal to a differential one. 

The Stanford-developed MCRC ASIC is designed to read out MIT-LL CCD detectors at the same speed and low noise performance of traditional discrete electronics, while consuming a fraction of the PCB footprint and power. Discussions of the design and performance results of the MCRC can be found in [\citenum{herrmann_mcrc_2020}], [\citenum{orel_x-ray_2022}], and [\citenum{{orel_x-ray_2024}}]. A newly designed daughter card that operates two parallel 8-channel MCRCs enables the simultaneous readout of up to 16 detector channels. 

In addition to source follower-mode p-channel junction field-effect transistor (pJFETs)\cite{bautz_toward_2018, bautz_progress_2020, bautz_performance_2022, chattopadhyay_development_2022}, we are also working on the development and characterization of the novel Single-electron Sensitive Readout (SiSeRO) devices to enable low-impedance drain current readout\cite{chattopadhyay_first_2022, chattopadhyay_improved_2023, chattopadhyay_demonstrating_2024}. These output stages have been utilized in previously characterized detectors, as well as our current generation of CCDs:

\begin{itemize}
    \item The MIT-LL CCID-100 is a 1440 x 1440 pixel charge-coupled device (CCD) detector with 16 parallel channels based on a pJFET output amplifier stage. Combined with two MCRC chips, this readout chain is the representative prototype of the AXIS focal plane flight hardware\cite{miller_high-speed_2023, donlon_directions_2024}. 
    \item The MIT-LL CCID-93++ detectors are a family of CCID-93 variants designed to test different configurations of output amplifier stage arrays. Each variant features 16 pJFET or SiSeRO based output stages, which can be read out in either parallel or in series\cite{donlon_directions_2024}. 
\end{itemize}
 
We continue to use an FPGA-based Semiconductor Technology Associates, Inc. Archon controller\footnote{http://www.sta-inc.net/archon/} to interface with our readout electronics. The Archon controller is a full data acquisition (DAQ) system that drives both the detector and the MCRC chips, while simultaneously capturing detector pixel data from both MCRCs in parallel. It has an internal digital signal processing engine to analyze and render the image frames. Furthermore, it is equipped with a Graphical User Interface (GUI) that enables interactive control of relevant signals including clocks and biases, as well as data-taking abilities for asynchronous analysis. A detailed description of our configuration of the Archon modules is provided here [\citenum{chattopadhyay_tiny-box_2020}].

Commissioning of the Archon CCD controller was performed using a previously characterized MIT-LL CCID-93 detector and pre-amplifier board. The baseline noise, gain, and spectral noise density performance results were used to verify the successful integration and denoising of the controller in a new lab environment.  

\section{X-Ray Fluorescence}
\label{sec:XRF}

In order to characterize the CCID-100 detectors across the wide energy band required by AXIS\cite{miller_high-speed_2023}, the beamline houses an X-ray fluorescence (XRF) setup capable of producing mono-energetic photon emission lines covering the 0.3-10 keV energy range of interest. Our setup functions by utilizing broadband bremsstrahlung radiation produced by an X-ray source to illuminate a wheel mounted with target materials. Characteristic X-ray photons are excited in the selected material as electrons transition into the innermost K-shell, which then travel along the beamline body toward the detector. The design of the source end is based on that of the Gen1.0 source system, which is described in [\citenum{stueber_xoc_2024}]. To improve line fluxes and spectral contamination, we have made several changes to the Gen2.0 design, which are reported below.  

\begin{figure} [ht]
\centering
\begin{subfigure}{0.55\textwidth}
    \includegraphics[width=\textwidth]{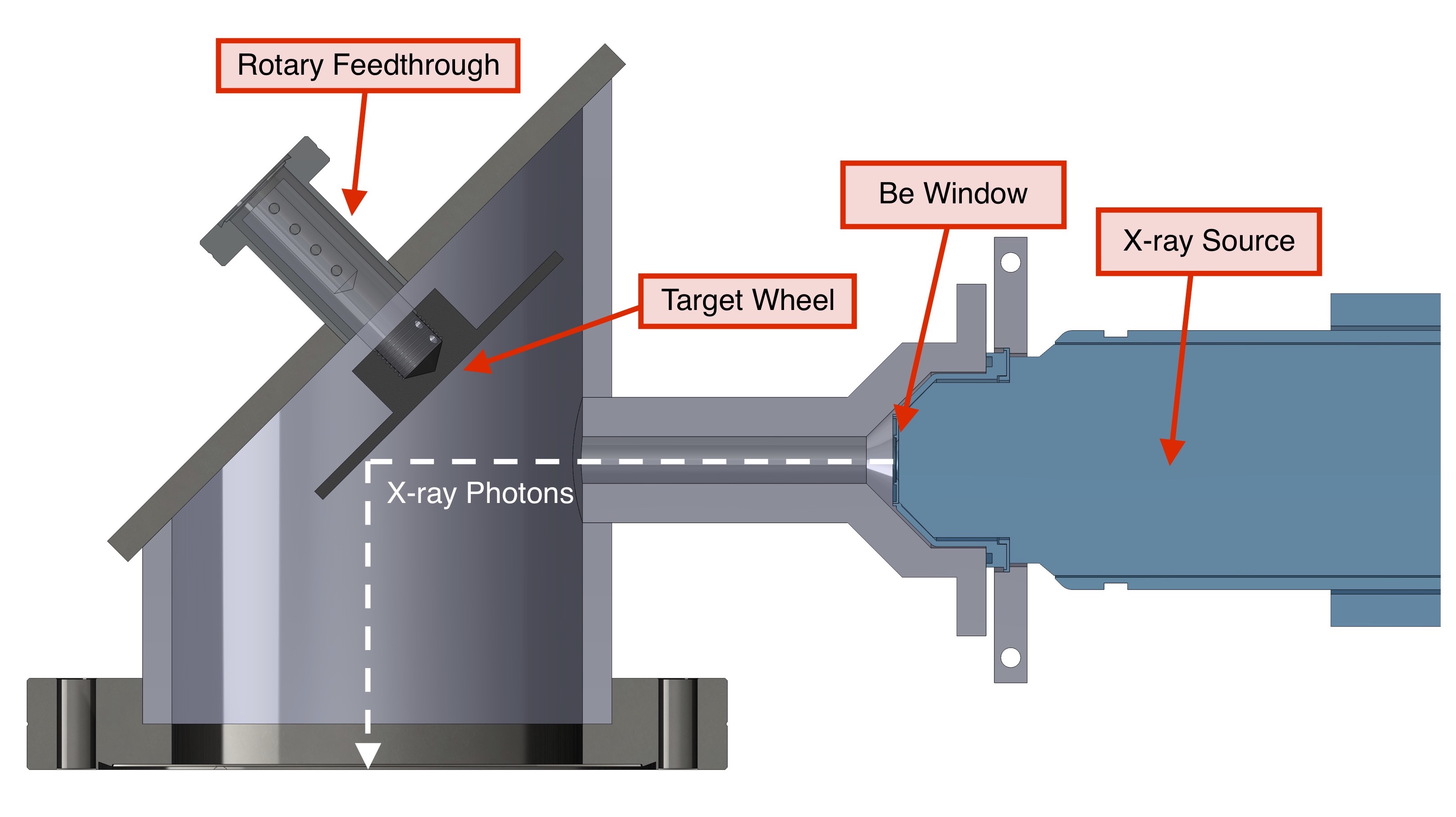}
    \caption{}
    \label{fig:XRFmod}
\end{subfigure}
\begin{subfigure}{0.33\textwidth}
    \includegraphics[width=\textwidth]{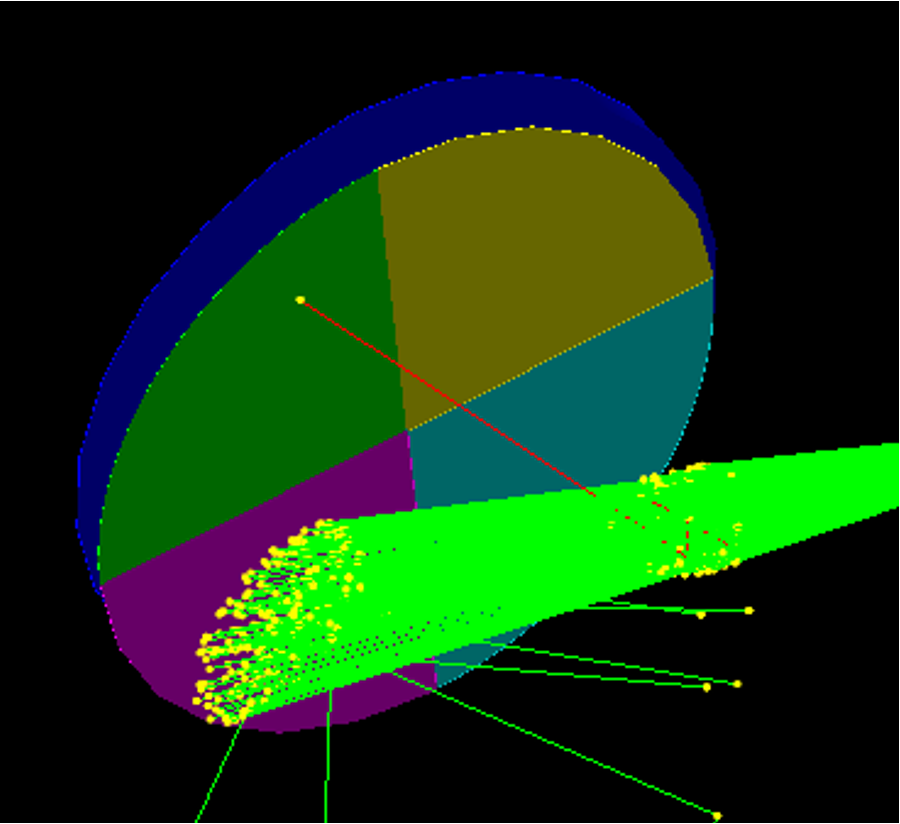}
    \caption{}
    \label{fig:spotsize}
\end{subfigure}
   \caption[sourcend] 
   { \label{fig:sourcend}(a) Solidworks model of the updated X-ray Fluorescence source end. Blue denotes the outline of the X-ray source. The source contains a 50 $\mu$m Be radiation window. A wheel with four target materials can be adjusted in-situ via a rotary feedthrough. The white dashed line indicates the travel path of photons emitted from the source and subsequently excited fluorescent photons from the target materials. (b) Geant4 simulation of the X-ray spot size on the target wheel after 1000 source photons. The beamline body has been removed for better visibility.}
   \end{figure}

\subsection{X-Ray Source}
\label{sec:Source}
To increase the flux of our low-energy targets, we use a Micro X-ray Lightbright End Window X-ray Tube\footnote{microxray.com/products/end-window-x-ray-tube/} as our primary X-ray source. The source generates X-rays by using high voltages to accelerate electrons from a heated cathode filament towards to tungsten anode. Energy generated in the anode is radiated as X-rays, or else released as heat\cite{anburajan_overview_2019}. The energy of the outgoing X-rays are set by the excitation voltage of the source, while the flux depends on the beam current of the anode \cite{bruker-axs_introduction_2008}.  

Compared to the side-illuminated tube utilized in the Gen1.0 beamline, which has a maximum beam current of 2.0 mA\cite{microx-ray_seeray_2022}, this end-illuminated source is rated to produce 11.0 mA when run at excitation voltages of 3kV to 9kV\cite{micro_x-ray_inc_lightbright_2023}, allowing for a higher emission rate of low-energy photons. In addition, a thinner 50 $\mu$m beryllium window reduces the attenuation at soft energies\cite{chantler_energy_1995}. Power is provided to the source by a Spellman uXHP 50 kV 100 W high voltage power supply\footnote{https://www.spellmanhv.com/en/high-voltage-power-supplies/uXHP} with an external interlock switch for user protection.  

\subsection{X-ray Fluorescence Assembly}
\label{sec:Body}

We have updated the design of the source end body to accommodate the different mounting requirements and beam spread of the end-illuminated X-ray source. The vacuum seal is maintained with a Viton o-ring\footnote{https://www.mcmaster.com/9464K301/}. We adjusted the length and inner diameter of the mounting flange to provide additional collimation of the X-ray beam and control the resulting illuminated spot size on the target wheel. The spot size was optimized to achieve the maximum illumination of the selected target material, while minimizing spectral contamination from adjacent targets or the stainless steel source end body. A Geant4 simulation of the expected spot size is shown in Fig. \ref{fig:spotsize}. 
   
As in our previous design\cite{stueber_xoc_2024}, we mount one of two rotatable target wheels into the source end. Both wheels hold four target materials that can produce fluorescence emission lines across the 0.3-10 keV energy range, such that the full energy range of interest can be produced on one wheel. The selected targets\footnote{Target materials purchased from Goodfellow Advanced Materials: https://www.goodfellow.com/} and their purity, thickness, and emission energies are summarized in Table \ref{tab:materials}. The targets are positioned with alternating high and low characteristic energies, to help with identifying potential spectral contamination if targets are misaligned during operation. We install a silicon drift detector (SDD)\footnote{Ametek XR123 25mm$^2$ Fast Silicon Drift Detector with C2 (silicon nitride) X-ray window: https://store.amptek.com/x-123-fast-sdd-with-low-energy-c-window-model-xr123-1-5-inch-extender/} above our test detector assembly that provides in-situ spectral data for checking the target wheel alignment \cite{stueber_xoc_2024}. We also operate the SDD monitor when characterizing test detectors to provide concurrent measurements of incident photon flux for quick estimates of detector quantum efficiency.

\begin{table}[ht]
\caption{Target Materials and Corresponding Thickness, Purity, and Emission Energies} 
\label{tab:materials}
\begin{center}       
\begin{tabular}{|c|c|c|c|} 
\hline 
\rule[-1ex]{0pt}{3.5ex} \textbf{Target Material} & \textbf{Thickness} & \textbf{Purity} & \textbf{Emission Energies} \\
\hline
\rule[-1ex]{0pt}{3.5ex} 
\multirow{2}{*}{Teflon (C$_2$F$_4$)}& 
\multirow{2}{*}{200$\mu$m}& 
\multirow{2}{*}{N/A} & 
0.27 keV (Carbon K$\alpha$)\\
&&&0.67 keV (Fluorine K$\alpha$)\\
\hline
\rule[-1ex]{0pt}{3.5ex} 
Graphene& 
1000$\mu$m& 
99.99\%& 
0.27 keV (Carbon K$\alpha$)\\
\hline
\rule[-1ex]{0pt}{3.5ex} 
Magnesium& 
100 $\mu$m& 
99.9\%& 
1.25 keV K$\alpha$\\
\hline
\rule[-1ex]{0pt}{3.5ex} 
\multirow{2}{*}{Quartz (SiO$_2$)}& 
\multirow{2}{*}{250 $\mu$m}& 
\multirow{2}{*}{N/A}& 
0.52 keV (Oxygen K$\alpha$)\\
&&&1.74 keV (Silicon K$\alpha$)\\
\hline
\rule[-1ex]{0pt}{3.5ex} 
Aluminum & 
200$\mu$m& 
99.999\%& 
1.48 keV K$\alpha$\\
\hline
\rule[-1ex]{0pt}{3.5ex} 
\multirow{2}{*}{Titanium }& 
\multirow{2}{*}{100 $\mu$m}& 
\multirow{2}{*}{99.99\%}& 
4.51 keV K$\alpha$\\
&&&4.93 keV K$\beta$\\

\hline
\rule[-1ex]{0pt}{3.5ex} 
\multirow{2}{*}{Iron}& 
\multirow{2}{*}{125 $\mu$m}& 
\multirow{2}{*}{99.99\%}& 
6.39 keV K$\alpha$\\
&&&7.06 keV K$\beta$\\
\hline
\rule[-1ex]{0pt}{3.5ex} 
\multirow{2}{*}{Copper}& 
\multirow{2}{*}{100 $\mu$m}& 
\multirow{2}{*}{99.99\%}& 
8.04 keV K$\alpha$\\
&&&8.91 keV K$\beta$\\
\hline
\end{tabular}
\end{center}
\end{table}


A gate valve in the beamline allows the source end to be removed for target wheel installation without venting the main beamline body. The connection to the source end is a standard 6'' outer diameter ConFlat vacuum flange\footnote{https://www.lesker.com/flanges/flanges-cf-304ss/part/f0600x000n}. This allows the beamline to be compatible with multiple source end designs and enables the system to be reconfigured to accommodate test setups in other energy regimes. 

\begin{figure} [ht]
\centering
\begin{subfigure}{0.4\textwidth}
    \includegraphics[width=
    \textwidth]{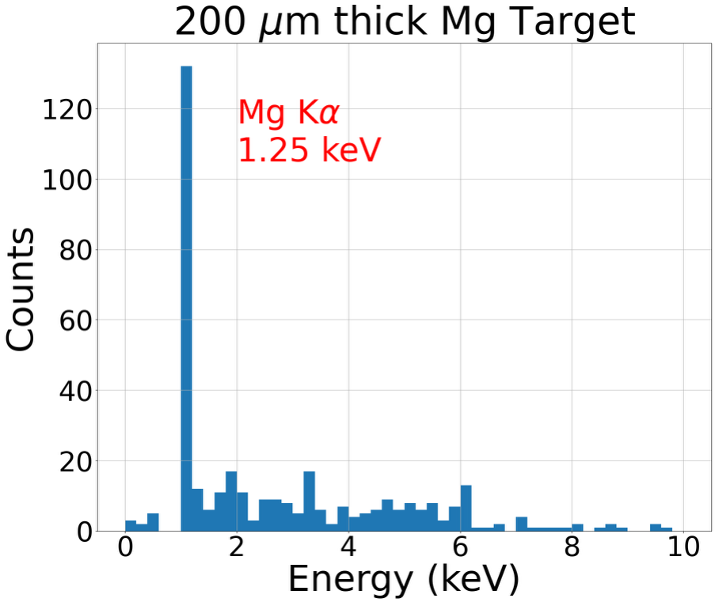}
    \caption{}
    \label{mg_spec}
\end{subfigure}
\hspace{1cm}
\begin{subfigure}{0.4\textwidth}
    \includegraphics[width=\textwidth]{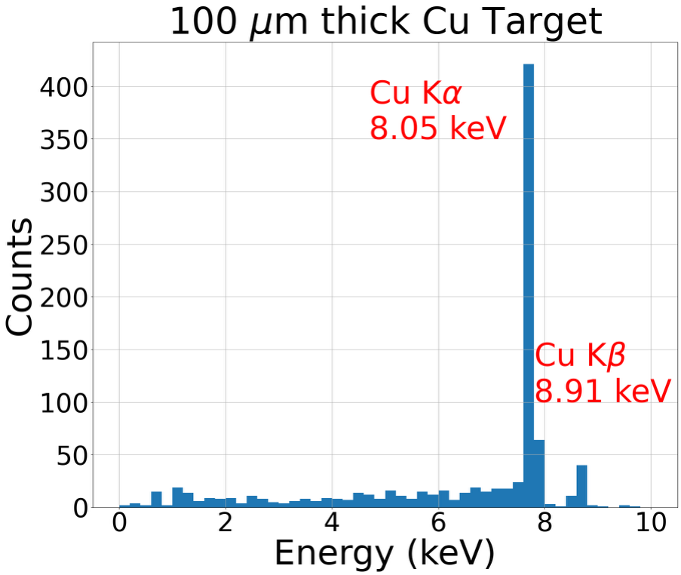}
    \caption{}
    \label{cu_spec}
\end{subfigure}

\caption[G4] 
{ \label{fig:G4} Simulated GEANT4 spectra for (a) a 200 $\mu$m thick Mg target and (b) a 100 $\mu$m thick Cu target. 50 million source photons were used in each simulation. }
\end{figure}

\subsection{GEANT-4 Simulations}
\label{sec:G4sim}
To verify the performance of the redesigned source end, we ran GEANT-4\footnote{https://geant4.web.cern.ch/} particle simulations to model the behavior of the XRF setup. Details of the model construction, material compositions, and interaction physics are reported by [\citenum{stueber_xoc_2024}]. In brief, the geometries of the beamline body, detector chamber, and source end body were defined using imported CAD models\cite{poole_cad_2012}. The detector and surrounding mechanical assembly (i.e. cooling block, thermal strap, clamping plates, standoffs) were defined with built-in GEANT4 shape functions. Smaller components such as screws, rods, and wires were assumed to have negligible impact and were excluded from the simulations. 

Some changes were made to the model to capture spectral contamination and flux losses at the source end. In addition to the XRF target material of interest, we modeled the fully mounted target wheel with all four materials, and verified that the illuminated spot did not overlap into adjacent quadrants (Fig. \ref{fig:spotsize}). The photon source was also redefined as a 35\textdegree{}  diverging beam to model the cone angle of the X-ray source\cite{micro_x-ray_inc_lightbright_2023}. 

Simulations with 50 million source photons were run for the target materials. Results for the magnesium and copper targets respectively are presented in Fig. \ref{fig:G4}, showing the number of events that interact with the detector volume in the 0-10keV energy band. We observed minimal contamination from adjacent sources compared to the characteristic K$\alpha$ and K$\beta$ emission lines of each target.

\section{Summary and Future Plans}
\label{sec: Summary}

The XOC Gen2.0 X-ray Beamline has been assembled and commissioned, with updated systems improving pumping efficiency, cooling performance, X-ray fluorescence flux at low energies, and spectral contamination. Along with a similar setup at MIT Kavli Institute, both XOC beamlines will work in parallel to enable a high volume of detector characterization efforts. In addition to the commissioning and characterization of the AXIS CCID-100 detectors, these beamlines will support the testing of the MIT-LL CCID-93++ prototype detectors, which include output arrays using p-JFETs and SiSeROs. The XOC beamlines are designed with the flexibility to accommodate a range of detector setups and energy bands in response to the technology gaps identified in NASA's most recent Prioritized Technology Gap Report. These twin systems are the foundation of a lab-scale testing facility capable of supporting current and future detector technology maturation efforts for future astronomical observatories.

\acknowledgments 
 
We are grateful to the Stanford School of Humanities and Sciences for funding to support laboratory and beamline commissioning. We also thank the Stanford Physics Machine Shop for modifying the detector chamber door and manufacturing the source-end body. AXIS detector characterization work is supported by the NASA APEX grant 80GSFC25CA019 and NASA SAT grant NSSC23K0211.        

\bibliography{references}
\bibliographystyle{spiebib} 

\end{document}